# Laser-driven Collisionless Shock Acceleration of Ions from Near-critical plasmas

S. Tochitsky[1*], A. Pak[2], F. Fiuza[3], D. Haberberger[4], N. Lemos[2], A. Link[2], D.H. Froula[4], and C. Joshi[1]

[1] Department of Electrical Engineering, UCLA, Los Angeles, California 90095, USA
[2] Lawrence Livermore National Laboratory, Livermore, CA, 94550, USA
[3] SLAC National Accelerator Laboratory, Menlo Park, CA 94025, USA
[4] Laboratory for Laser Energetics, University of Rochester, 250 East River Road, Rochester, New York 14623-1299, USA
*Invited Speaker. e-mail: sergei12@ucla.edu  Paper NI3.6 Bull. Am. Phys. Soc. **64**, 176 (2019)

This paper overviews experimental and numerical results on acceleration of narrow energy spread ion beams by an electrostatic collisionless shockwave driven by 1 μm (Omega EP) and 10 μm (UCLA Neptune Laboratory) lasers in near critical density CH and He plasmas, respectively. Shock waves in CH targets produced high-energy ~50 MeV protons (ΔE/E of ≤ 30%) and 314 MeV $C^{6+}$ ions (ΔE/E of ≤ 10%). Observation of acceleration of both protons and carbon ions to similar velocities is consistent with reflection of particles off the moving potential of a shock front. For shocks driven by $CO_2$ laser in a gas jet, ~30 MeV peak in He ion spectrum was detected. Particle-in-cell simulations indicate that regardless of the target further control over its density profile is needed for optimization of accelerated ion beams in part of energy spread, yield and maximum kinetic energy.

I. **Introduction**
Collisionless shocks produced by colliding plasmas are present in a variety of astrophysical settings and are responsible for accelerating very high energy particle beams in space e.g. shocks in Supernova Remnants are considered to be a dominant mechanism in cosmic ray generation[1-3]. Electrostatic shock related mechanisms for acceleration of ions in the context of space and laboratory plasmas have been investigated in the 60s and 70s both theoretically and numerically[4-6]. The key finding of these studies is that in a collisonless hot plasma, reflection of ions off a moving shock front is the main dissipation mechanism necessary for solitary propagation of electrostatic shock with a velocity $v_{sh}=MC_s$, where M is the Mach number and $C_s$ is ion sound speed. Here, an electrostatic potential eΦ associated with a moving shock wave can effectively reflect ions with the kinetic energy smaller than that of the electrostatic potential[4]. As was shown numerically by Forslund and Freidberg[5], reflection of a small fraction of the upstream ions converts a well known ion acoustic wave into a structure with a steep potential gradient upstream and with damped downstream oscillations behind the front. If the velocity of the shock is quasi constant, ions reflected off this moving shock front can be accelerated into a narrow quasi-monoenergetic distribution at a velocity ~$2v_{sh}$. Such Collisionless Shock Acceleration (CSA) of ions is interesting not only from a basic plasma physics point of view but also as a mechanism for generating energetic ion beams with a narrow energy spread, once conditions for shock excitation and its laminar propagation with a constant speed are possible to reach in a laboratory plasma.

With an advent of high-peak-power lasers it was realized that a focused laser beam can be used to locally compress the plasma due to radiation pressure and simultaneously heat up the plasma electrons thus creating conditions for a shock formation in a laboratory laser-plasma[7-9].

However, as was shown in 2D PIC simulations for foils with a sharp rear edge, a sheath field of this plasma-vacuum boundary causes smearing of energy spectra of ions leaving the target and results in production of a plateau extending to high-energy side of a characteristic to Target Normal Sheath Acceleration (TNSA) continuous exponential decreasing energy spectra [9,10].

It is known that interactions of high-intensity lasers with solid-density targets can accelerate ion beams to 10s of MeV/u by a well-studied TNSA mechanism (see e.g. a review by Macchi et al[11] and references therein), or TNSA enhanced via radiation induced transparency[12,13] or so-called break-out afterburner mechanisms[14]. Such accelerated beams are of interest for a broad range of potential applications[11,15,16]. Until today, however production of high-energy and high-quality: low-energy spread and focusable ion beams from a laser-plasma interaction still remains a challenge. Therefore exist an interest to the alternative acceleration mechanisms that can generate ion beams with a narrow energy spread and CSA is potentially one of them.

Recent proof-of-existence experiments have demonstrated that monenergetic proton beams with energies up to 22 MeV can be generated by interaction of a multi-terawatt, picosecond $CO_2$ laser pulse ($\lambda \sim 10$ μm) with a $H_2$ gas jet[17]. Our interferometry measurements have uniquely related these accelerated proton beams to a near-critical density tailored plasma profile. It has a steepened front with a rise around $10\lambda$ created by the laser radiation pressure on a critical density layer and a smooth exponentially falling ramp on the back with a characteristic length of $30\lambda$. Detailed 2D PIC simulations for such a tailored plasma profile revealed that monoenergetic protons are accelerated by a collisionless shock structure formed at the near critical density layer and propagating at $v_{sh} \sim 0.16c$. Such high shock velocity results from significant heating of plasma electrons to relativistic electron temperatures $T_e \geq 1$ MeV[17,18]. The exponential long ramp on the rear side of the plasma profile allowed for the extraction of shock reflected monoenergetic protons because the TNSA field was significantly reduced and constant[19,20]. While using long-wavelength lasers for ion acceleration from gas targets is promising due to possible operation at a high repetition rate, relatively low charge and limited $CO_2$ laser power stimulated us to study more developed and powerful 1-μm high-power lasers for CSA of ions.

In this paper we summarize a nearly decade long studies at the Neptune Laboratory (UCLA), Jupiter Laser (LLNL), and Omega EP Laser (LLE) facilities on laser-driven CSA of ions in a near critical density plasma with the tailored plasma profile and associated numerical simulations. The paper is organized as following. In Sec. II we briefly describe the basic dynamics of CSA. Using particle in cell (PIC) code simulations we identify signatures of CSA from mixed species CH plasmas. In Sec. III we present a summary of the experimental observations obtained in CSA experiments with picosecond 1-μm lasers at Omega EP. Here, in a two-beam configuration a laser produced x-ray source causes ablation of the back of the target as well as reducing the peak plasma density due to hydrodynamic expansion. Once the peak plasma density $n_e$, drops close to relativistically corrected critical plasma density, $n_e \sim a_0 n_c$, (where $a_0 = eA/m_e c^2 \geq 1$ is a normalized laser vector potential), a high-power short pulse driver (laser piston) strikes this preformed plasma and launches a shock wave. Ions reflected off the moving shock front are accelerated to high energies via the CSA mechanism are recoded and analyzed. In Sec. IV we summarize our two-beam experiments conducted with a high-intensity Titan 1 μm laser, and discuss similarities in physics of CSA in multispecies CH plasmas In Sec. V we extend $CO_2$ laser gas jet interactions study to other than $H_2$ gases and describe laser-driven CSA of He ions by 10 μm picosecond pulses. We also discuss potential of a tunable H, He, N, O…Ne ion source based on CSA in a gaseous plasma which is desirable for medical applications[21]. Thus in this paper we demonstrate that in both 1 µm laser-foil and 10 µm laser-gas interactions electrostatic collisionless shocks can generate high-energy ion beams with a narrow energy spread. Furthermore, we find that regardless of the laser wavelength and the target material control over the plasma density profile is critical for optimization of accelerated ion beam properties such as energy spread, yield and maximum kinetic energy.

## II. Collisionless shock ion acceleration in tailored near critical density CH plasmas using 1 µm lasers

The ability to utilize the high intensities (>$10^{21}$ W/cm$^2$) of widely available solid state 1 µm glass laser systems to drive collisionless shock waves is appealing as the shock velocity, and hence reflected particle velocity scales with the plasma electron temperature. Assuming pondermotive heating of electrons ($T_e \propto a_0 \propto I^{0.5}$), the reflected ion velocity scales with the laser intensity by approximately $2v_{sh} \propto I^{0.25}$. However, creating an idealized plasma density profile suitable for driving a shock, while preserving the narrow energy spread of the reflected ion, as described in our previous papers[17,18,20] is challenging.

There are three main considerations for such targets. The first is to have the peak plasma density be of order the relativistic critical density, $n_e = a_0 n_c$. Here the critical density is given by $n_c = \frac{m_e \varepsilon_o \omega_0^2}{q_e^2}$ and for 1 µm wavelength lasers is 1.1x$10^{21}$ cm$^{-3}$. Additionally, a small density scale length on the leading edge of the target, such that the plasma density rises as quickly as possible, allowing the laser to efficiently compress and heat the plasma in order to create the pressure discontinuity required to launch the shock is also desirable. The second consideration is that the thickness of the target should be limited to $L = \lambda \left(\frac{m_i}{m_e}\right)^{1/2}$ to achieve a uniform electron temperature and shock velocity throughout the target. This sets a thickness requirement on the order of 50-100 µm for 1µm lasers, depending on the target material used. The third consideration is, that in order to maintain the narrow energy spread on the reflected ion beam, the density on the rear side of the target needs to decrease gradually in order to mitigate gradients in the sheath field of the target that broaden the energy spread of the reflected ion beam. As discussed theoretically[18,20], an ideal rear density profile is an exponential with a scale length $L_{g0} \approx \frac{\lambda_o}{2}\left(\frac{m_i}{m_e}\right)^{1/2}$. Such a rear density ramp creates a sheath field with a nearly uniform electric field that continues to accelerate the shock reflected ions without smearing the energy distribution. As will be further discussed, the final accelerated energy of the shock reflected ions is strongly influenced by the density profile.

While previous work was conducted using a single species hydrogen plasma, in experiments with 1 µm lasers, multi-species CH plasmas are used. Therefore, to understand the dynamics of collisionless shocks in multi-species CH plasmas, PIC simulations were first conducted. To better understand the dynamics of ion acceleration by a laser-driven shock in a multiple ion species plasma, 2D PIC simulations with the code OSIRIS[22] were performed. The simulations modeled the interaction of a 1 ps duration 1-µm laser pulse with a peak $a_0$ = 8.5, with a CH plasma. The initial plasma has T$_e$= 5keV and a profile with a peak density of $10^{22}$ cm$^3$ ($n_e \approx a_0 n_c$), with a 10 µm linear ramp on the front side and an exponential ramp on the rear side with a scale length L$_p$ = 20 µm. In order to be able to simulate the long time dynamics of the interaction and given the associated computational constrains, we used a long and narrow simulation box, with 830 µm in the laser propagation direction and 10 µm in the transverse direction. The simulation resolved this domain with 32768 x 384 cells, used a time step t = 0.058fs, 16 particles/species/cell, cubic particle shapes, and ran for ~4 ps. The boundary conditions were open/absorbing in the longitudinal direction and periodic transversely.

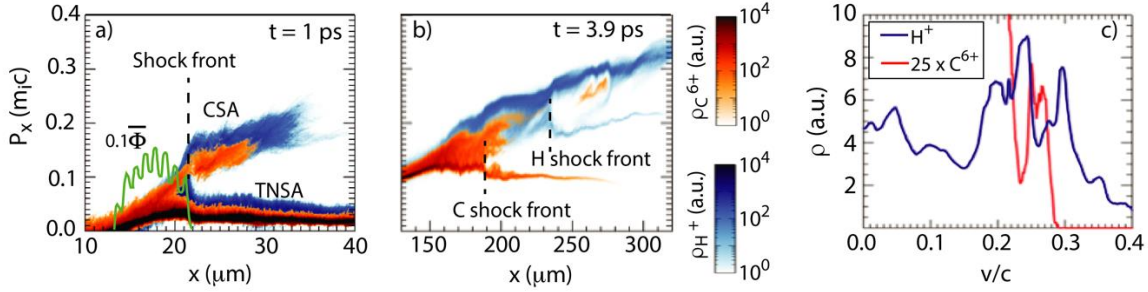

**Figure 1**: (Color) Ion phase space taken at 1 ps and 3.86 ps from the start of the ultra-intense short pulse laser in (a) and (b), respectively. In (a), the solid green curve denotes the potential associated with the electrostatic shock wave normalized by the kinetic energy of the upstream inflowing protons. For ease of display, the normalized potential has been multiplied by 0.1 and indicates that $H^+$ ions will be reflected at a value of 0.1 while the 2× lower charge to mass $C^{6+}$ ions will be reflected at a value of 0.2. Integrated velocity spectrum of hydrogen ions, shown in blue and $C^{6+}$ ions shown in red, at 3.86 ps from the start of the short pulse laser (c).

We note that the 2D simulation presented here have focused on describing the role of the longitudinal plasma profile and composition on CSA and due to the computational expense used a periodic simulation box transversely with a plane-wave like laser profile. The results of 2D simulations with finite laser spot size have been shown and discussed in ref.18,20 where for the parameters studied they yielded similar results to plane-wave simulations. While these results are encouraging, we should note that 3D simulations capturing the full multi-dimensional effects for the extended plasma profiles considered in these studies are extremely demanding and have not yet been reported.

The simulations show the formation of an electrostatic shock and reveal that in this collisionless multiple ion species plasma there is a strong species separation between $H^+$ and $C^{6+}$ ions, due to the different A/Z. This leads to the development of a two-step electrostatic potential associated with the compression of the two ion species, which modifies the spatial and temporal dynamics of the ion reflection. Figures 1a and 1b show the temporal evolution of the $H^+$ and $C^{6+}$ phase space. Simulations indicate that the shock is formed behind the peak density in the bulk of the plasma, and begins to reflect both $H^+$ and $C^{6+}$ during the laser irradiation of the target. After ~1 ps, the shock potential has fallen below that required to reflect $C^{6+}$, but remains large enough to reflect $H^+$ (Fig. 1 a). At 3.86 ps in Fig. 1b, we observe that part of the upstream protons are reflected at the hydrogen shock front (x ~ 235 μm) and part is reflected at the carbon shock front (x ~190 μm). This leads to two peaks in the proton spectrum and a single peak at the $C^{6+}$, as shown in Figure 1c. Given that $C^{6+}$ are only reflected at early times, their number is considerably smaller than protons. The ion spectrum is no longer evolving significantly by the end of the simulation. The final velocity of the reflected ions ($v_f$) is given by the sum of the contributions from the TNSA field and the shock reflection. Specifically, $v_f = 2v_{sh}^* + v_{TNSA}$, where $v_{TNSA}$ is the ion velocity due to the TNSA field and $v_{sh}^*$ is the shock velocity in the frame of the upstream fluid, i.e. $v_{sh}^* = v_{sh} - v_{TNSA}$. We observe that $v_{sh}^* \sim 0.06c$ through most of the acceleration process and $v_{TNSA,H+} \sim 2v_{TNSA,c6+}$. This is consistent with acceleration over a fixed time corresponding to the decay of a uniform TNSA electric field associated with the exponential plasma profile[23]. Based on these observations, we can estimate the relationship between the velocity of the different reflected ion species as,

$$\frac{v_{f,H^+}}{v_{f,c^{6+}}} = \frac{2v_{sh}^* + v_{TNSA,H^+}}{2v_{sh}^* + v_{TNSA,H^+}/2} \quad (1)$$

We can immediately see that if shock reflection dominates $v_{f,H^+}/v_{f,c^{6+}} \sim 1$, whereas in the limit where the TNSA dominates $v_{f,H^+}/v_{f,c^{6+}} \sim 2$. In the simulation we have $v_{f,H^+}/v_{f,c^{6+}} \sim 1.16$. In the experiments discussed in previously published studies[24] $v_{f,H^+}/v_{f,c^{6+}} \sim 1 - 1.3$. This strongly suggests the role of shocks in these CSA experiments.

Thus these simulations indicate that in a multi-species CH plasma, the signatures of CSA include the generation multi peaked ion spectra and the acceleration of species with different charge to mass ratios to similar velocities. These signatures were observed in experiments conducted using the 1 µm Omega EP laser at LLE as well as at Titan laser at LLNL and the main findings from the latter were discussed in detail elsewhere[24]. In Sec. IV we highlight the physics in two-beam CSA experiments using multiple Titan laser shots. But first, in the next section, we describe in detail experiments on CSA of ions, using the most powerful~0.5 PW, 1 µm laser pulses at the Omega EP facility. Note that in recent experiments where two co-propagating 1 µm laser beams were used to accelerate ions in near critical density[25] and underdense[26] CH plasmas, no proton beams with the narrow energy spread were detected as one may expect based on the absence of a controlled pre-expanded rear-side plasma. The observed high energy tail in the ion distribution was explained by the CSA enhancement inferred from PIC simulations.

## III. Experimental results on CSA of ions at Omega EP

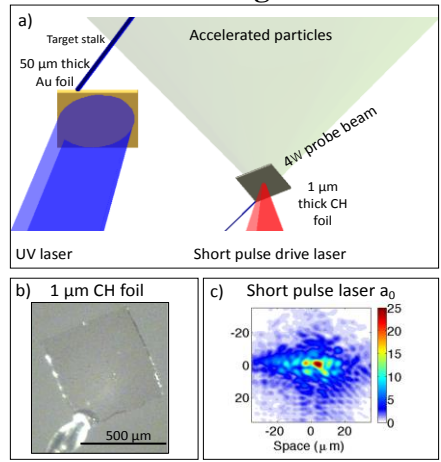

Figure 2: Experimental setup at Omega EP. A 1 ns duration UV laser first irradiates a Au foil to create an x-ray source that ablates initially 1 μm thick CH foil. The expansion of the CH foil is measured by the 4ω probe beam. After waiting 0.3 to 0.5 ns, a short pulse drive laser is used to create the shock (a). Photograph of initial CH target (b). Inferred $a_0$ of the short pulse drive laser which is used to create the shock c).

At Omega EP, a laser-produced x-ray source was used to create a plasma density profile suitable for CSA. As seen in Fig. 2a, the x-ray source was produced by using a 351 nm UV laser, with a 1 ns long square pulse duration, focused onto a 50 μm thick gold foil. The focused spot diameter was 750 μm and the average intensity across the spot was $2.8 \times 10^{14}$ W/cm$^2$. At these conditions, the radiation hydrodynamic code HYDRA[27] was used to estimate the radiative flux produced. These calculations indicated a spectrum that was fit well to that of a blackbody emission with a kT~165 eV. The center of the 1 mm x 1 mm gold foil was positioned 3.37 mm from a CH target foil. An image of the CH target foil is shown in Fig. 2b. The CH foil was initially 500 μm x 500 μm wide and had a thickness of 1 μm. The ablation of the rear side of the CH foil creates a gradually decreasing density profile required to reduce the energy broadening of the shock reflected particles when leaving the plasma. By adjusting the delay between the UV laser and the high intensity 1μm laser used to drive the shock, the peak plasma density could be tuned to match the laser intensity to optimize coupling and shock formation. After waiting 0.3 to 0.5 ns for the plasma to expand, the short pulse drive 1μm laser was used to irradiate the CH target and drive a shock. The drive laser had a pulse duration of ~750 fs and a total energy up to ~500 J. An example of the inferred on-shot $a_0$ distribution for the drive laser is shown in Fig. 2c and indicates that a peak $a_0$ of ~25 was achieved. The calculated average $a_0$ within the central 10 μm radius was 8.6. Accelerated ions were diagnosed using a magnetic spectrometer with a slit width of 250 μm in the deflection direction, resulting in a ΔE/E uncertainty of 2% at 50 MeV. In the non-dispersed direction, the slit width was 5 mm long. We note that for TNSA shots, always taken with a single 1 μm beam at half of the laser power, a smooth monotonically decreasing spectrum was observed.

For the two-beam configuration, as seen in Fig. 2a, the expansion of the CH foil was measured using angular filter refractometry (AFR) using the 4ω (263 nm) probe beam[28]. An example of the AFR signal 0.58 ns after the UV laser irradiates the gold foil is shown in Fig. 3a. For this shot, the data quality was not impacted by the short pulse drive laser as only the UV laser was used, and the axis of the target was well aligned to the 4ω probe laser axis. In order to estimate the plasma density profile, radiation hydrodynamic simulations using HYDRA[27] were first used to generate the expected x-ray flux incident on the CH target foil produced by the UV irradiation of the gold foil. Simulations then used this x-ray flux to irradiate a 1 μm CH target. The resulting plasma density profiles and synthetic AFR images were then generated. Figure 3b shows a comparison between the measured AFR profile (blue line) and the best fit to that profile obtained from the HYDRA simulations (red line). Figure 3b shows that the overall extent of the refracted light is best fit by a synthetic AFR image that is generated ~80 ps after the measured temporal delay. In general, the agreement between the synthetic AFR and the measured AFR are in fair agreement. However, the offset in timing, as well differences in the number of AFR fringes suggest that the expansion is slightly faster than predicted and that gradients in the actual plasma density profile are likely not as steep as those calculated. Figure 3c shows the predicted temporal dependence of the plasma density profile. Here the peak density is observed to decrease from 33.2 to 13.4 $n_c$ as the delay between the UV and short pulse laser is increased from 0.3 to 0.5 ns.

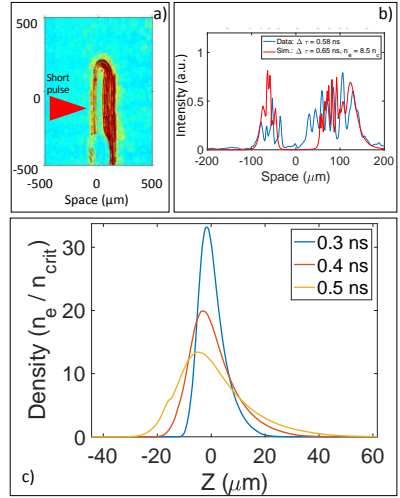

Figure 3: Side on AFR image taken 0.58 ns after the start of the 1 ns UV laser pulse (a). The expansion is observed to be asymmetric, with the x-rays preferentially heating and ablating the rear surface of the foil. Comparison of the measured AFR profile to a synthetic AFR profile produced from ray tracing through a simulated density profile (b). Simulated electron density profiles for 3 different expansion times showing density of between 13 and 33 $n_c$ can be achieved for delays between 0.3 and 0.5 ns (c).

Figure 4 a shows the raw dispersed ion signal recorded on the image plate detector. The signal was obtained for a temporal delay between the UV pulse and short 1 μm drive pulse of 0.35 ns. Using HYDRA calculations this delay corresponds to a density profile with a peak $n_e$ of $25n_c$. However as discussed, the most accurate measurements of the foil expansion indicated a faster expansion of the foil than calculated. Taking this into account, the peak plasma electron density is estimated to be ~$17n_c$. The laser $a_0$ for this experiment is shown in Fig. 2c. As seen in Fig. 4a, differential filtering of 50 and 300 μm of aluminum was applied to the image plate detector to discriminate between the lighter hydrogen and heavier carbon ions. The 50 μm thick aluminum filter transmitted all protons over 2.5 MeV and all $C^{6+}$ ions above 46 MeV, while the thicker filter was 300 μm thick and transmitted all protons over 6 MeV and all $C^{6+}$ ions over 146 MeV. The raw dispersed spectrum shows an exponentially decreasing component with a sharp cutoff around 2.7 cm. Beyond this a spectrally narrow, double peaked distribution of $H^+$ ions are observed. As seen in Fig. 4b, the two $H^+$ peaks are located at ~45 and 54 MeV, respectively. The overall full width half maximum of this distribution is 20 MeV. Both the shape and spectral width of the double peaked $H^+$ ion spectrum is consistent with expectations of shock accelerated spectra details in Fig. 1. Figure 4a shows that at a position of 1.3 cm there is a narrow spectral feature transmitted through both filters. A slight decrease in the amplitude across the filter boundary was consistent with the signal being $C^{6+}$ ions at an energy of 314 MeV. The spectral width of this $C^{6+}$ feature is 32 MeV. Figure 4b shows the velocity of both the $H^+$ and $C^{6+}$ ion distributions are comparable with the observed ratio of $v_{f,H^+}/v_{f,C^{6+}} \sim 1.35$, which is consistent with what is expected for accelerated shock accelerated distributions of ions and what has been also previously observed in Titan experiments[24]. Note that Thomson parabola data (not shown) recorded with an ~1 mm diameter pinhole confirmed that besides $C^{6+}$ no other carbon ions were detected above the noise.

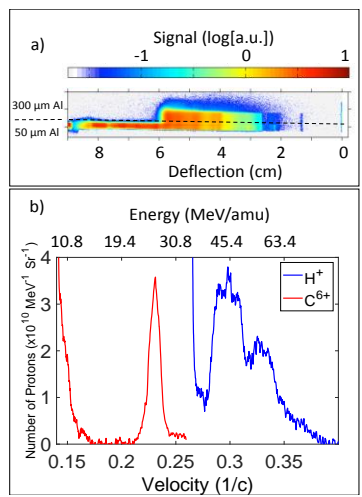

Figure 4: Dispersed ion spectrum observed for a 0.35 ns delay between the UV and short pulse drive lasers (a). The dashed line indicates the regions of 50 and 300 μm aluminum filtration, used to isolate the $H^+$ from the C ion signals. The resulting $H^+$ and $C^{6+}$ ion spectrum, showing narrow distributions of comparable velocities, indicative of shock acceleration (b).

The energy of the spectrally narrow $H^+$ peaks is 2.5-3X higher than that observed during the experiments conducted at the Titan laser. Additionally, the number of protons /MeV/Sr was observed to be ~10X larger than ~$3 \times 10^9$ particles measured in previous experiments[24]. The high peak energy and larger spectral intensity are generally expected, as the experiments at Omega EP used a laser with a 2.8X larger $a_0$, a spot size with ~4X the area and a target with a peak plasma density that was 1.4-2X larger than that used in the Titan experiments. Simulations suggest that both the maximum peak energy and number of shock accelerated ions can be further optimized using the Omega EP laser and the plasma density profiles that can be achieved via x-ray ablation of thin foils. However, a limited number of shots allowed at this facility prevented careful optimization of the plasma profile. The sensitivity to the plasma density profile and reproducibility of SWA ions was studied using a larger number of shots undertaken using a lower power 1 μm Titan laser pulses ($a_0 \leq 9$).

**IV. Summary of the CSA experiments at the Titan Laser Facility**
A series of two-counterpropagating beam experiments on ion acceleration were performed at the Titan Laser Facility at LLNL. In these experiments, a 0.5 μm CH target (half the thickness of the Omega EP experiments discussed above) was first irradiated from the back side by a 10 ns long, 1 μm ablation laser at an average peak intensity of ~$10^{11}$ W/cm$^2$. After the target expansion, an ~1 ps laser driver ($a_0 \leq 9$) was sent to the tailored plasma at a variable delay. In Fig.5 we present multiple proton spectra recorded for nearly the same laser power and different delay time controlling the peak plasma density of the target at the arrival of the laser piston-drive pulse. Without the ablation laser (Fig.5a), the TNSA spectrum extends to 19 MeV. At a 3 ns delay (in Fig. 5b), HYDRA calculations estimate the peak electron density $n_e > a_o n_c$ at the driver pulse and the cut-off energy decreases to 11 MeV due to a reduced sheath field. Apparently at an optimal delay of ~4 ns, the spectrum in Fig. 5c shows a narrow distribution of protons stretching up to 20 MeV and energy density in the peak above that of the TNSA shot. Slight drop in the drive laser power in Fig. 5d resulted in a proton peak to be around 15 MeV. At this delay/plasma conditions spectra with narrow energy spread peaks were consistently observed at energies between 10-20 MeV. Specifically, we observed the proton energy of the highest energy peak can vary between 11.4 and 17.7 MeV. This represents a ~45% shot-to-shot variation in the energy of the ion spectral peaks. This variation, even at nominally similar laser conditions, indicates that the overall laser-plasma coupling and shock dynamics varied from shot to shot (e.g. due to self-focusing, thermal lensing of the laser, higher spatial modes of the laser profile and ~20 μm accuracy to which we can position the target shot to shot). Differential filtering of the magnetic spectrometer data as well as Thomson parabola spectra also exhibited $C^{6+}$ ion beams propagating with a similar velocity[24]. For longer delays, the peak plasma density $n_e < a_o n_c$, as predicted by HYDRA, and no protons above 3 MeV were observed (see Fig.5e).

Thus, systematic observation in two-beam experiments of multiple species of ions with different charge-to-mass ratios being accelerated to similar velocities and into narrow spectral peaks is consistent with the reflection and acceleration off a moving shock front potential. In addition, measurements of electron temperature presented in Fig.5f indicate that the number and slope temperature of hot electrons in the optimal near critical density plasma profile is much higher than that measured in the regular TNSA case. The efficient coupling to hot electrons is critical to drive a fast electrostatic shock wave with a modest Mach number 1.5-2 capable of accelerating ions to high kinetic energies as it propagates in the laser heated plasma[18,20]. We observe that the higher electron temperature corresponds to the higher maximum energy of the ion spectral peak.

In these experiments, the measured ion velocity was 2.5X lower than that calculated using PIC simulations with an idealized profile (see Fig.1). As discussed by A. Pak et al,[24] current understanding is that in an experimentally produced laser ablated profile, the rear side plasma density falls faster and extends to longer distances than in the idealized exponential profile. Modeling for the experimental ablated plasma profile resulted in a lower final kinetic energy of 23 MeV with a ΔE/E of 34% which agrees rather well with experimental observations of narrow distributions of protons between 11.4-17.7 MeV with ΔE/E of ~10-20%.

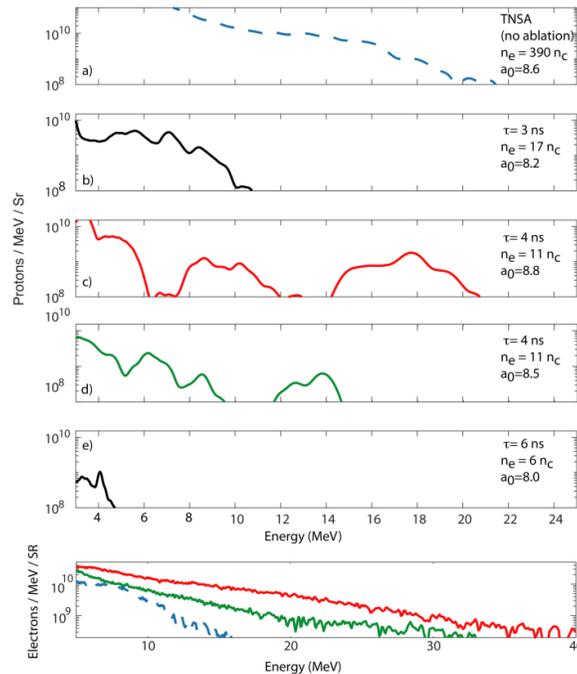

**Figure 5**. Spectra of accelerated protons (a-e) measured for different time delays between the ablation and driver laser pulses: (a) for a nonablated foil; (b-e) for delays in the range of 3-6 ns. Electron spectra recorded for some of the above shots are presented in (f).

**V. He ion shock acceleration in a gas jet plasma using picosecond $CO_2$ laser**

Several science and medical applications would benefit from a relatively low cost source of light ion beams. In particular, He, N, C, O, Ne ion beams are ultimately considered to be the optimal tool to treat advanced hypoxic and radiation resistant tumors because of broader Bragg peaks, than that of protons, and still insignificant ion splitting (the smallest is for He ions)[21,29]. Their use is currently hindered by existence of only few and very expensive carbon ion therapy facilities and as a result scant data on relative biological effectiveness (RBE) for other light ions from helium to neon exists.

The demonstration of acceleration of high-energy proton beams with a narrow energy spread of $\leq 10\%$ using $CO_2$ laser-driven electrostatic collisionless shocks in a gas jet plasma introduced a potentially transformative concept in the laser-driven ion acceleration research[17,18]. The use of a gas target instead of a foil is very attractive because it is a clean source of protons and other ions from He to Ne. It can be run at a high-repetition rate, and the density of the plasma can be adjusted around the critical plasma density, $n_c \sim 10^{19}$ cm$^{-3}$ for a 10 μm pulse. It should be noted that several groups have been pursuing idea of CSA ion acceleration in a gas jet not only using 10 μm[30] but 1 μm lasers[31-33] as well. However, interaction of ~1 μm laser with $H_2$ gas jet generated protons with a smaller energy in the 1-6 MeV range[31-33] indicating challenges in the target control for $n_e > 10^{21}$ cm$^{-3}$ density. In this section we describe our results on He ion acceleration to an energy of ~30 MeV by the CSA mechanism at 10 μm.

The experiment was carried out at the UCLA Neptune Laboratory that houses a multi-TW $CO_2$ laser system[34]. In this study a ~10 μm laser beam with an energy up to 50 J was focused by an F/3 parabolic mirror onto a 1.4-mm diameter He gas jet. The peak neutral density of gas was $2\text{-}3\times10^{19}$ cm$^{-3}$. The energy of laser pulse was distributed over an ~100 ps long pulse train of ~3 ps micropulses separated by 18.5 ps[34]. The maximum laser intensity reached $2.5\times10^{16}$ W/cm$^2$, corresponding to $a_0=1.4$ for the highest intensity micropulse. The accelerated ions in a forward direction were intercepted by a stack of CR-39 nuclear track detectors with a thickness of 300 and 1 mm that are not sensitive to X-rays. By analyzing the penetration depth of He ions using a calculated Bragg peak value for each energy, we were able to recover the spectrum of particles. Note that no dispersive devices were used, so for the following we assumed that He$^+$ ions dominated the particle beams coming from the gas jet target despite the fact that He$^{2+}$ ionization threshold could be reached locally in a plasma. It is mainly related to charge exchange/recombination of ions propagating through a layer of partially ionized or neutral helium atoms in the downstream part of the gas jet. Note that this assumption also correlates with the peak plasma density measurements described below. Fig.6a depicts a He ion spectra recorded for a 46J $CO_2$ laser pulse when the laser focus was moved upstream by a distance of about the Rayleigh range from the center of a gas jet. A narrow energy peak was detected around 30 MeV in addition to low energy ions. When the laser was focused in the middle of the gas jet, for the same detector stack only low-energy ions from 0.2-1 MeV were recorded above the noise level related to mechanical defects of CR-39 plastic. Within a 50x50 mm$^2$ detector, the ion yield in a narrow high energy peak was measured to be ~$5\times10^5$. Finally, it is important to note that in earlier detailed optimization of CSA of protons from a $H_2$ gas jet[17] the upstream position of the laser focus was also optimal for production of high-energy proton beams. Apparently minimization of the laser beam break-up and electron filamentation in the plasma were achieved in the experiment by propagating a slightly diverging laser beam (the laser focus upstream) in the plasma in order to compensate for self-focusing.

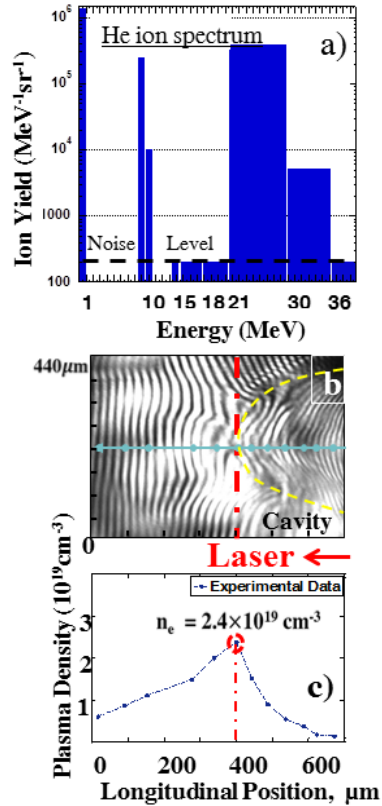

**Figure 6**. Experimentally recorded He ion spectra for a $CO_2$ laser beam focused to an intensity of ~$2.5\times10^{16}$ W/cm$^2$ in a gas jet (a). An interferogram taken at the peak of the laser macropulse (b) and corresponding extracted plasma density profile on-axis of the laser shown by a blue dots (c). The measured uncertainty in plasma density is about 10%.

$CO_2$ laser-plasma interactions in the gas jet occur both in underdense and overdense plasmas. Few of the main processes are laser filamentation and nonlinear self-focusing in underdense (transparent) part of the ramped plasma, a classical hole boring or pushing of the steepened density plasma by the radiation pressure at a critical density layer and transport of a stream of hot electrons from the overdense (opaque) plasma through the neutral gas. To gain an insight into such complicated dynamics, the evolution of plasma density profile during the $CO_2$ laser pulse train was monitored by a four-frame interferometry using 1 ps, 532 nm pulses[35]. A typical He plasma profile taken right after the maximum micropulse hit the plasma is presented in Fig.6b. Here the earlier micropulses have already produced a parabolic shaped cavity in plasma (indicated by a dashed yellow line). Outside the cavity there is a higher density ~50 μm thick plasma wall where some blurring of the fringes has occurred due to strong refraction of the probe beam. Therefore, a point by point manual analysis of the interferogram has been performed to extract the on-axis plasma density[35]. It is shown by a blue line in Fig. 6b. Extracted plasma density values along this line are presented in Fig. 6c. Here the measured peak plasma density is $2.5\times10^{19}$ cm$^{-3}$ and apparently the plasma profile has a steepened front and a smooth long scale back sides. Although the laser could not penetrate the overcritical plasma layer, photoionization by bremsstrahlung radiation and collisional ionization by electrons accelerated by laser are thought to be responsible for a millimeter scale exponentially falling plasma density formed on the back of the gas jet.[17] The good agreement between the measured values of plasma density and the initial neutral gas density indicates single ionization of He atoms in the overall volume of plasma. This supports the $He^+$ ion spectrum in Fig.6a and suggests that $He^{2+}$ may locally be produced within the filamented laser beam but it was not measurable by the interferometry diagnostic.

Using a fiducial placed slightly off the plane of the gas jet, we were able to carefully measure the relative displacement of the maximum density layer for the four time frames within a ~100 ps time window corresponding to the $CO_2$ laser macropulse. Such direct single-shot measurements of the laser hole boring revealed that the maximum hole boring velocity, $v_{hb}$ reached ~7 x$10^{-3}$c at the peak of the laser pulse train. An electrostatic field associated with a moving layer of high-density plasma can reflect ions and they will gain maximum energy of $2v_{hb}$. Such reflection off the moving overdense plasma layer could result in acceleration of ions to ~0.4 MeV much smaller than the observed energies. This so called hole boring radiation pressure acceleration[11,36] therefore can not be responsible for high-energy ion peak shown in Fig. 6a.

The mechanism of high-energy He ion peak observed in our experiment was revealed using 2D PIC code OSIRIS. It was found that, similar to hydrogen plasmas[17], reflection of He ions from the electrostatic field potential of a moving collisionless shock wave is responsible for acceleration. The laser pulse is absorbed/reflected near the critical density layer, heating up the plasma and forming locally a density spike due to radiation pressure. Both an electron temperature gradient and plasma density discontinuity causes formation of an electrostatic shock [5,18,20]. Once formed, the shock propagates through the plasma at nearly constant velocity of $v_{sh}$ for almost 100 ps after the laser pulse. Here the laser-driven collisionless shock wave is propagating in the near critical density plasma with a velocity $v_{sh}$ and is reflecting ions from the background plasma to a velocity of $v_{ions}$~$2v_{sh}$. As the shock overtakes the plasma ions, they can be reflected off it and gain a velocity twice the shock velocity minus initial ion velocity. This process forms a helium ion beam that retains its narrow energy spread as it exits plasma. The latter is achieved due to a long scale (100s of microns) exponential plasma ramp, in which the shock experiences constant and very small sheath fields[18]. For $a_0 \geq 1$, the energy of $He^+$ ions reached 30 MeV. Simulations also revealed importance of launching a quasi-1D shock for generating a forward propagating ion beam with a small divergence. This supports the experimental finding that a slightly diverging laser beam, used in the case of upstream focusing, was optimal for producing forward propagating high-energy He ions (see Fig.6a).

Using 2D PIC modeling we have also studied scaling of energy of He ion beams to kinetic energies around 200 MeV. In Fig. 7 for $a_0$=5, $He^{2+}$ ions are reaching velocities of 0.3c which corresponds to an energy of 170 MeV nearing the requirements for radiotherapy. Such laser field strength corresponds to $3 \times 10^{17}$ W/$cm^2$ intensity at 10 μm and can be reached with $\geq$25 TW pulses which are possible to achieve by upgrade of an existing picosecond $CO_2$ laser system[34].

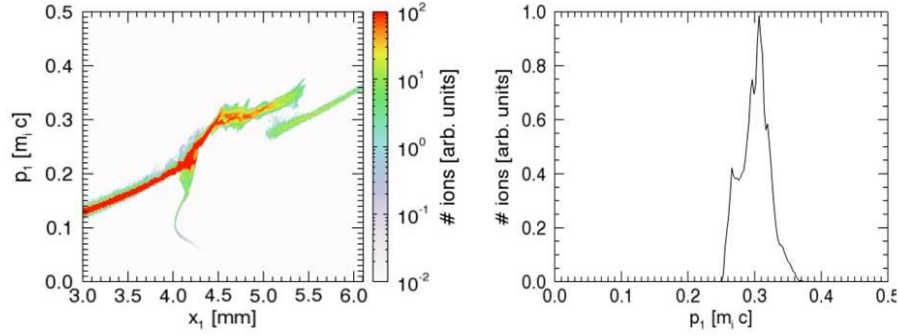

**FIGURE 7.** Phase space of $He^{2+}$ ions and spectrum of the ion bunch in SWA at the $CO_2$ laser intensity of $3 \times 10^{17}$ $W/cm^2$.

As mentioned above, many applications would benefit from a 1-10 Hz picosecond MeV ion source easily tunable from protons to He or…Ne light ions. At these repetition rates, a 3ps long 10 μm pulse focused to an intensity of $10^{15}$-$10^{16}$ $W/cm^2$ can be used for SWA of ions from a gas jet. We have analyzed numerically the minimum laser power required for driving a collisionless shock in a near critical density plasma. At an intensity I=$3 \times 10^{15}$ $W/cm^2$ ($a_0$=0.5), shock reflected few percent of protons and accelerated them to 3.5MeV. The accelerated beam with an energy spread ≤10 % then can be focused by a pulsed solenoid[37]. Such focused beam could be either used for RBE measurements in biological samples or could be coupled into a RF cavity synchronized with the $CO_2$ laser in order to decrease the energy spread of the beam to better than 1% as described by A. Noda et al[38]. It should be noted that in order to minimize effect of laser-plasma instabilities and have flexibility in choosing a charge state of accelerated ions we also consider using an intense focused 0.8 μm laser beam for ionization of a ~20-30 μm diameter plasma column. Then a quasi-1D shock can be launched in a preformed ~$10^{19}$ $cm^{-3}$ plasma channel by a high-power 10 μm pulse. An important issue for a high-repetition rate ion source is development of a small, 100 μm-scale diameter gas jet which should minimize the charge exchange effect and increase the ion yield.

## VI. Conclusions

In summary, we show in this paper that laser-driven SWA in a laboratory plasma can be used for producing high-energy ion beams with a narrow energy spread. Regardless of type of the target or the laser wavelength, we found that for producing high-energy ions it is important to launch a shock wave in a relativistic hot plasma ($T_e$>1 MeV) which can be accomplished in laser-plasma interaction at a near critical peak plasma density.

In the case of 1 μm laser driven shocks in a near critical density CH plasma, a new two-beam configuration is applied for tailoring the plasma profile. In the Omega EP experiment, ablation plasma on a rear side of a 1 μm thick CH foil is produced by x-ray radiation before a short pulse intense 1 μm laser piston drives a collisionless shock wave in a preformed plasma. By controlling the plasma density profile via tuning the time delay between these two beams, at optimal delays we detected high-energy ~50 MeV protons (ΔE/E of ≥30%) and 314 MeV $C^{6+}$ ions (ΔE/E of ≥10%). Observation of acceleration of both protons and carbon ions to similar velocities and detection of multiple ion peaks in a spectra are experimental evidences of particle reflection off the moving field front associated with the shock wave. Note that similar ion spectra were reported in our previous experiments at Titan[24]. Results from PIC simulations identify the CSA mechanism to be responsible for ion acceleration in multiple species plasma. Modeling also indicates importance of further control of the plasma profile in order to achieve even higher energies (without extra laser power) and higher ion yields. We believe that a two-beam configuration gives a limited ability for such control of the plasma profile and consider a specially designed target with a smoothly falling density gradient on the back of a CH foil for future single 1 μm laser beam CSA experiments.

In the case of $CO_2$ laser driven shocks in He gas jet, ions were accelerated to ~30 MeV indicating potential of developing a tunable light ion source at a high repetition rate. The ion yields detected so far are low and we have several hypothesis ranging from the hot electron-driven shock forming in the low density upstream plasma to accelerated He ions recombining on the way to the detector due to charge exchange. Extra studies are needed to understand the physics behind the observations and to optimize the gas jet target. Such short light ion bunches can then be further accelerated in a cyclotron to the requisite energy for ion-therapy or can be used as a laboratory ion source for RBE studies.

**Acknowledgements:**
S.T. would like to thank Luis O. Silva (IST, Portugal) for initial interest and support and stimulating discussions over many years in this scientific endeavor. This work was performed under the auspices of the U.S. Department of Energy (DOE) under Contract No. DE-AC52-07NA27344, with support from the LLNL Laboratory Directed Research and Development Program under tracking code 15-LW-095. The work at UCLA was supported by NNSA Grants No. DE-NA0003842 and DE-NA0003873, and DOE grant DE-SC0010064. Experiments were enabled at Omega EP under the DOE sponsored Laboratory Basic Science program. D. Haberberger and D. Froula contributions were supported by the U.S. Department of Energy under Cooperative Agreement No. DE-NA0001944.

**Data availability statement:**
The data that supports the findings of this study are available from the corresponding author upon reasonable request.